\documentclass[RNAAS]{aastex631}

\usepackage{graphicx}
\RequirePackage[fixed]{fontawesome5}
\newcommand{\github}[1]{%
   \href{#1}{\faGithub}%
}
\usepackage{natbib}
\usepackage{hyperref}
\usepackage{float}

\begin{document}

\title{ELMA: ELlipse-based bar MAjor axis estimator}

\correspondingauthor{Bruna R. Bragança de Lima}
\email{bruna.rosa@inf.ufrgs.br}
\author[0009-0002-4295-741X]{Bruna R. Bragança de Lima}
\affiliation{Instituto de Informática, Universidade Federal do Rio Grande do Sul, 91501-970, Porto Alegre, RS, Brazil}
\affiliation{Instituto de Física, Universidade Federal do Rio Grande do Sul,  91501-970, Porto Alegre, RS, Brazil}

\correspondingauthor{Andressa Wille}
\email{andressa.wille@ufrgs.br}  
\author[0000-0001-6687-0402]{Andressa Wille}
\affiliation{Instituto de Física, Universidade Federal do Rio Grande do Sul,  91501-970, Porto Alegre, RS, Brazil}

\correspondingauthor{Rafael S. de Souza}
\email{rd23aag@herts.ac.uk, drsouza@ad.unc.edu}
\author[0000-0001-7207-4584]{Rafael S. de Souza}
\affiliation{Centre for Astrophysics Research, University of Hertfordshire, College Lane, Hatfield, AL10~9AB, UK}
\affiliation{Department of Physics \& Astronomy, University of North Carolina at Chapel Hill, NC 27599-3255, USA}
\affiliation{Instituto de Física, Universidade Federal do Rio Grande do Sul,  91501-970, Porto Alegre, RS, Brazil}

\author[0000-0000-0000-0002]{Ana L. Chies-Santos} 
\affiliation{Instituto de Física, Universidade Federal do Rio Grande do Sul,  91501-970, Porto Alegre, RS, Brazil}

\begin{abstract}
Galactic bars are key non-axisymmetric structures in disk galaxies, driving angular-momentum redistribution and contributing to secular evolution, central mass build-up, and the formation of nuclear structures. Robust and homogeneous measurements of bar length, however, remain challenging, particularly for large imaging surveys, where manual estimates are time-consuming and sensitive to methodological choices. We introduce \textsc{elma}, a standalone, pip-installable Python package for automated bar-length estimation in galaxies already identified as candidate barred systems. The method operates directly on two-dimensional imaging data, using iterative elliptical-isophote fitting to trace the radial ellipticity profile and identify a projected bar-length estimate from the semi-major axis associated with the local maximum in ellipticity. Using the image WCS information and a user-supplied redshift, \textsc{elma} converts angular measurement into a projected physical length. We demonstrate the package on JWST/NIRCam imaging of barred galaxies in GOODS--South field. The code is released under the MIT license at \href{https://github.com/BrunaLimaa/elma}{this repository \faGithub}.
\end{abstract}


\section{Introduction}

Galactic bars are among the most prominent non-axisymmetric structures in disc galaxies, redistributing angular momentum and serving as tracers of secular dynamical evolution \citep[e.g.,][]{sheth2005, debattista2006}. Measuring these structures across large-scale surveys requires automated and reproducible methods that reduce human subjectivity. This need is amplified by the current and forthcoming generation of photometric surveys, which will provide high-quality imaging for millions of galaxies, demanding efficient pipelines for homogeneous morphological measurements and catalogue construction.

Recently, the study of galactic bars has entered a new observational regime with the capabilities of the James Webb Space Telescope (JWST), which has enabled the detection of bars at high redshifts ($z > 2$), challenging current theoretical models of disk settling and stability \citep{costantin2023, conte2024}. At the same time, understanding the internal dynamics that lead to the formation of nuclear structures, such as nuclear discs and rings, requires precise and systematic morphological decomposition. 

Despite their prevalence, bar-length measurements remain sensitive to the adopted detection and measurement methodology, affecting reported bar fractions and scaling relations.
Ellipse fitting is a well-established technique for characterizing bars and estimating their radial extent \citep{erwin2005, marinova2007, aguerri2009}, but its application has traditionally required substantial manual supervision, creating a bottleneck for large-scale studies. Complementary approaches include citizen-science measurements of bar sizes \citep{hoyle2011, hutchinsonsmith2026} and recent deep-learning methods that derive bar lengths from segmentation \citep{walmsley2023, cavanagh2024}. These developments highlight both the scientific demand for scalable bar measurements and the continuing need for transparent, reproducible tools.

In this context, we present \texttt{elma}, a scalable and easy-to-use framework for automated bar-length estimation in galaxies already identified as barred or as candidate barred systems. \texttt{elma} uses iterative isophote fitting to measure the radial extent of the bar-dominated region directly from two-dimensional imaging data. For a user-provided target redshift, the projected physical bar length is calculated by scaling the measured angular size by the angular-diameter distance, $D_A(z)$. We assume a $\Lambda$CDM cosmology with $H_0 = 70~\mathrm{km~s^{-1}~Mpc^{-1}}$ and $\Omega_m = 0.3$.

\section{Methodology}

The \texttt{elma} pipeline estimates galaxy bar lengths through a three-stage workflow: data preparation, iterative isophote fitting, and cosmological size calibration. First, the pipeline reads the input FITS file and extracts the image data and the World Coordinate System (WCS) information required for angular and physical size conversion. For two-dimensional images, the surface-brightness map is used directly. For three-dimensional data cubes, the cube is collapsed along the spectral axis by summation, producing a two-dimensional morphological map.

Second, the galaxy centre is defined as the position of the brightest pixel. Starting from this centre, \texttt{elma} uses the \texttt{photutils} package \citep{photutils} to perform iterative elliptical-isophote fitting following the method of \citet{jedrzejewski1987}. 
At each semi-major axis, the surface-brightness distribution along a trial ellipse is sampled as a function of the eccentric anomaly $E$ and represented by a harmonic expansion,
\begin{equation}
I(E) = I_0 + \sum_{n=1}^{N}\left[A_n \sin(nE) + B_n \cos(nE)\right],
\end{equation}
where $I_0$ is the mean intensity along the isophote and $A_n$ and $B_n$ quantify deviations from a purely elliptical contour. In the iterative fitting procedure, the low-order harmonic terms are used to update the eccentricity ($\epsilon$), and position angle until convergence.
The procedure is initialized with a generic structural guess, adopting a semi-major axis of $a=5$ pixels and  $\epsilon=0.2$, and then proceeds outwards with a fixed linear step size of 0.05 pixels. This initialization avoids the innermost PSF-dominated region, reducing the impact of central noise. 
The code fits the central component on the assumption that this is bar-dominated, and its extent is then defined as the valid isophote corresponding to the local maximum in $\epsilon$. 

This quantity should be interpreted as a projected bar-length estimate, defined by the semi-major axis of the maximum-ellipticity isophote, rather than as a fully deprojected or dynamical bar radius. No inclination correction is applied, and \texttt{elma} does not perform a bulge--bar--disc decomposition; prominent bulges or other overlapping structures may therefore affect the ellipticity maximum used to define the estimate.
This angular size is then converted into a projected physical length in kpc using the angular-diameter distance, $D_A(z)$, evaluated at the user-provided redshift under the adopted flat $\Lambda$CDM cosmology.

\section{Analysis}

\begin{figure}[h]
\centering
\includegraphics[width=0.7\textwidth]{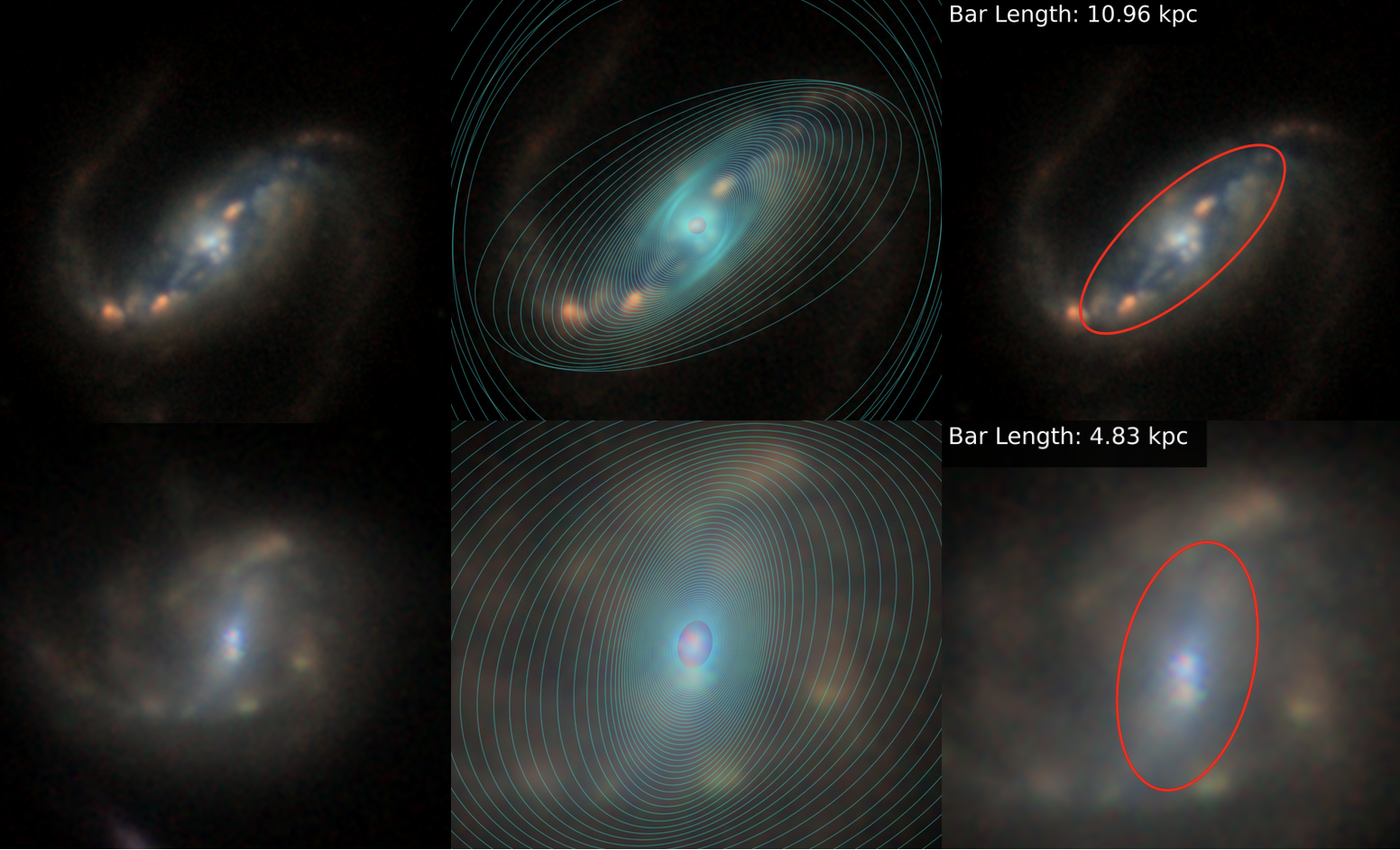}
\caption{Example \texttt{elma} outputs for two barred galaxies in the GOODS--South field.  Upper row: strong barred galaxy at $z=0.42$ (RA = 03:32:39.27, Dec = $-27$:45:32.97). Lower row: barred galaxy at $z=0.62$ (RA = 03:32:41.42, Dec = $-27$:46:51.72). Left: input RGB image. Middle: all fitted isophotes from the iterative ellipse-fitting procedure. Right: bar-dominated region enclosed by the best-fitting isophote associated with the local maximum in ellipticity. The measured bar extent is converted into a physical length using WCS information and galaxy redshift.}
\label{fig:diagnostic}
\end{figure}

We demonstrate the method using two JWST-observed barred galaxies from the GOODS--South field in the Advanced Deep Extragalactic Survey \citep[JADES;][]{rieke2023, eisenstein2025, eisenstein2026}; a typical \textsc{elma} output is shown in Figure \ref{fig:diagnostic}. The automated isophote fitting traces the visible bar-dominated structure, successfully isolating the bar region at the point of maximum ellipticity (highlighted by the red ellipse). 
By integrating this geometric fit with the corresponding WCS metadata and standard cosmological parameters, the package seamlessly converts the pixel-based measurement into a physical scale, eliminating the need for manual integration steps.

\section{Conclusions}

\textsc{elma} provides a user-friendly implementation of automated bar-length estimation from galaxy imaging data. Designed for galaxies already classified as barred, the pipeline identifies the central region, assuming bar dominance, from the ellipticity profile and converts the measured angular scale into a projected physical length using WCS information and redshift. The resulting measurements can be used to construct homogeneous bar-length catalogues for large galaxy samples.

\begin{acknowledgments}
R.S.S. acknowledges support from the Conselho Nacional de Desenvolvimento Científico e Tecnológico (CNPq, Brazil, grants 446508/2024-1 and 315026/2025-1). 
ACS acknowledges support from FAPERGS (grants 23/2551-0001832-2 and 24/2551-0001548-5), CNPq (grants 312940/2025-4, 445231/2024-6, and 404233/2024-4), and CAPES (grant 88887.004427/2024-00). 
\end{acknowledgments}

\vspace{1em}
\software{\texttt{astropy} \citep{astropy}, \texttt{numpy} \citep{numpy}, \texttt{scipy} \citep{scipy}, \texttt{matplotlib} \citep{matplotlib}, \texttt{photutils} \citep{photutils}.}

%\newpage
\bibliography{references}
\bibliographystyle{aasjournal}

\end{document}